\documentclass[prl,twocolumn,floats,floatfix,aps,superscriptaddress,showpacs]{revtex4-1}

\usepackage{amsmath}
\usepackage{graphicx}
\usepackage{amsfonts}
\usepackage{amssymb}
\usepackage{color}
\usepackage{footnote}

\usepackage{bm}

\newcommand{\mZ}{\mathcal{Z}}
\newcommand{\mD}{\mathcal{D}}
\newcommand{\mT}{\mathcal{T}}
\newcommand{\mC}{\mathcal{C}}

\newcommand{\mL}{\mathcal{L}}

\newcommand{\be}{\begin{equation}}
\newcommand{\ee}{\end{equation}}

\newcommand{\bea}{\begin{eqnarray}}
\newcommand{\eea}{\end{eqnarray}}

\newcommand{\Tr}{\textrm{Tr}}
\newcommand{\md}{\mu_{\textrm{d}}}
\newcommand{\ms}{\mu_{\textrm{s}}}
\newcommand{\fei}{f^{\textrm{ext}}_i}
\newcommand{\Ted}{T_{\textrm{Ed}}}
\newcommand{\Bed}{\beta_{\textrm{Ed}}}

\newcommand{\flM}{f_{ _{\lambda_\textrm{max}}}}
\newcommand{\lM}{\lambda_\textrm{max}}

\DeclareMathOperator{\sgn}{sgn}

\begin{document}
 
\title{Edwards thermodynamics for a driven athermal system with dry friction}

\author{Giacomo Gradenigo}
\affiliation{Universit\'e Grenoble Alpes, LIPHY, F-38000 Grenoble, France}
\affiliation{CNRS, LIPHY, F-38000 Grenoble, France}

\author{Ezequiel E.~Ferrero}
\affiliation{Universit\'e Grenoble Alpes, LIPHY, F-38000 Grenoble, France}
\affiliation{CNRS, LIPHY, F-38000 Grenoble, France}

\author{Eric Bertin}
\affiliation{Universit\'e Grenoble Alpes, LIPHY, F-38000 Grenoble, France}
\affiliation{CNRS, LIPHY, F-38000 Grenoble, France}

\author{Jean-Louis Barrat}
\affiliation{Universit\'e Grenoble Alpes, LIPHY, F-38000 Grenoble, France}
\affiliation{CNRS, LIPHY, F-38000 Grenoble, France}

\date{\today}

\begin{abstract}
We obtain, using semi-analytical transfer operator techniques, the
Edwards thermodynamics of a one-dimensional model of blocks connected
by harmonic springs and subjected to dry friction.  The theory is able
to reproduce the linear divergence of the correlation length as a
function of energy density observed in direct numerical simulations of
the model under tapping dynamics.  We further characterize
analytically this divergence using a Gaussian approximation for the
distribution of mechanically stable configurations, and show that it
is related to the existence of a peculiar infinite temperature
critical point.
\end{abstract}

\maketitle 


Although systems governed by dissipative interactions do not obey
equilibrium statistical mechanics, there have been several attempts to
describe such systems with \emph{effective} equilibrium-like theories.
A paradigmatic example is the problem of amorphous packings of
frictional grains, for which an effective thermodynamic was proposed
by Edwards and coworkers~\cite{EO89,ME89,EM94,EG98,BHDC15}.  This
approach relies on the basic assumption that all mechanically stable
packings of grains occupying the same volume have the same
probability.  This is expected if the system is repeatedly perturbed
with ``extensive operations''~\cite{EM94}, like a shaking of the
grains followed by a fast relaxation to a blocked (mechanically
stable) configuration.  One can then build an effective thermodynamics
by determining all mechanically stable configurations (MSCs) of the
grains, and computing the mean values of physical observables from
flat averages over accessible blocked configurations.  The predicted
mean values can then be compared to dynamical averages obtained from a
given ``tapping'' protocol which samples blocked configurations.

For athermal systems in which an energy is defined, Edwards'
prescription can be formulated as follows \cite{BKVS00}.  One
postulates the existence of an effective temperature $\Ted=\Bed^{-1}$
such that the probability of a blocked configuration $\mC$ of energy
$E(\mC)$ takes the form \be P(\mC) = \frac{1}{\mathcal{Z}}\, e^{-\Bed E(\mC)}\,
\mathcal{F}(\mC),
\label{eq:edmes}
\ee where $\mathcal{Z}$ is a generalized partition function;
$\mathcal{F}(\mC)=1$ if $\mC$ is a MSC and $\mathcal{F}(\mC)=0$
otherwise (only blocked configurations have a nonzero probability).
This constraint is non-Hamiltonian, in the sense that it gives a zero
probability to (mechanically unstable) configurations having a finite
energy, whereas they would have a finite probability in canonical
equilibrium.  At first sight, Eq.~(\ref{eq:edmes}) looks like a
harmless generalization of equilibrium statistical mechanics, by
simply restricting the set of accessible configurations.  For
instance, introducing an upper bound $|x_i| < X_{\rm max}$ on harmonic
oscillators $x_i$ does not deeply affect their statistical properties.
However, the nontrivial point is that the constraints arising from MSC
are often much more complex than simply introducing a bound on
individual variables, in particular the constraint of mechanical
stability may itself introduce strong correlations in the system.
Notice that, beyond volume and energy, one may also take into account
other quantities when building an Edwards-type
thermodynamics~\cite{HHC07,HC09,BE09,BJE12,BZBC13} (e.g., the stress
tensor).

Several attempts have been done to test the Edwards scenario, not only
in packings of grains,
experimentally~\cite{NKBJN98,SGS05,LCDB06,NRRCD09} and
numerically~\cite{KM02,M04,MD05,PCN06,BK15}, but also, in abstract
models like spin and lattice
gas~\cite{BPS00,LD01,L02,BFS02,DGL02,DGL03}, and in glass and
spin-glass models~\cite{CN00,BKVS00,BKVS01,DL01,LD03}.  Typically, one
uses a specific tapping protocol to sample blocked states, and
compares the dynamical average of the observables to the thermodynamic
averages obtained from Eq.~(\ref{eq:edmes}).  Although it has been
shown explicitly in some cases that Edwards approach is not exact
\cite{DGL02,DGL03}, it is generally believed to be a reasonably good
description in many cases~\cite{BHDC15}.  The main difficulty with the
Edwards measure Eq.~(\ref{eq:edmes}) is that the partition function
$Z$, from which all thermodynamic quantities can be derived, is very
complicated to compute due to the complexity of the function
$\mathcal{F}(\mC)$ characterizing blocked states
\cite{BE03,BE06,BSWM08,WSJM11,APF14}.  Standard approaches are then
either to consider abstract
models~\cite{BPS00,LD01,L02,BFS02,DGL02,DGL03}, which are far from any
realistic system but simple enough to allow for an explicit solution,
or to resort to mean-field~\cite{SL03} or more involved~\cite{BE03}
approximations, which still capture part of the interesting
phenomenology, but (at least partly) miss relevant information about
spatial correlations in the system.

In this Letter, we introduce a realistic model in which Edwards thermodynamics can be computed exactly.
We investigate a one-dimensional model of frictional blocks connected by harmonic springs, subjected to a tapping dynamics.
Due to the one-dimensional geometry, statistical properties can be computed semi-analytically in the thermodynamic limit using
a transfer operator method.
Our numerical simulation and theoretical results lead both to an infinite temperature critical point, with a correlation length
diverging linearly with the stored energy density --a directly measurable quantities in numerical simulations.
We analytically confirm these results using a Gaussian approximation for the joint probability distribution of spring elongations,
and further characterize this critical point in terms of the divergence of energy and length fluctuations.

\paragraph{Simulations--}
Our model is represented by a one-dimensional 
chain of massive blocks connected by $N$ harmonic springs sliding on a
horizontal plane~\cite{BK67,CL89,GB10,BPG11,BGP14}.
Each particle is subjected to dry (Coulomb) friction.
The position of the $i^{\rm th}$-mass is denoted as $x_i$.
The effect of dry friction is twofold: when a block is sliding it is subjected to a dissipative
force proportional to the dynamic friction coefficient,
$f_{i,\textrm{diss}} = -\md mg \sgn(\dot{x}_i)$; when at rest,
in order to make it move one has to apply a force larger than the
static friction $|f_i|> mg \mu_{\textrm{s}}$.
We denote $\xi_i = x_i-x_{i-1}-l_0$ the elongation of the spring
connecting block $i$ and block $i-1$; $l_0$ is the rest length of the spring.
Taking into account an external force $\fei$ the total force exerted on a block is
$f_i = k (\xi_{i+1}-\xi_i) + \fei$, with $k$ the spring stiffness.  
We write the equation of motion in dimensionless form using the variables $\tilde{t}=t/\tau_{\textrm{0}}$,
$\tilde{x}=x/(g\tau_{\textrm{0}}^2)$, $\tilde{\fei} = \fei/(mg)$ and
$\tilde{l}_0=l_0/(g\tau_{\textrm{0}}^2)$, with $\tau_{\textrm{0}}=\sqrt{k/m}$.
We have run simulations of a chain with $N+1=256$ blocks with open boundary
conditions -- see Supplementary Material (SM).
We checked that, over the range of forces explored, no finite size effects appear changing the size from $64$ to $2048$ blocks.
Dropping the tildes, the dimensionless equation of motion reads
\be\label{eqofmotion}
\ddot{x}_i = -\md \sgn(\dot{x}_i) + x_{i+1}+x_{i-1}-2x_i + \fei ,
\ee
while the condition to start motion becomes simply
$|\xi_{i+1}-\xi_i +\fei| > \ms$.
We identify the ``blocked'' configurations as those that
in absence of external force are mechanically stable: $\forall~i$,
$\dot{x}_i=0$ and $|\xi_{i+1}-\xi_i|<\ms$. 
We then define the following \emph{tapping} dynamics:
the external forces $\fei$ are switched on in Eq.(\ref{eqofmotion}) and act during a given period of time $\tau$, after which
they are switched off and the system relaxes to a MSC.
This procedure, that we call \emph{driving cycle}, is repeated a large number of times to sample MSCs.
At each cycle, the forces $\fei$ are drawn (randomly for each site $i$) from a distribution
\be
p(\fei) = (1-\rho)\, \delta(\fei) + \frac{\rho}{\sqrt{2\pi \sigma^2}}\, e^{-(\fei-F)^2/2\sigma^2}
\ee
A driving protocol is determined by fixing the parameters $\rho$ and $\sigma$.
For a given protocol, one can then vary the intensity $F$ and duration $\tau$ of the driving.
Each MSC is characterized by the typical value of the energy stored by the springs
$e=(1/2N)\sum_{i=1}^N\xi_i^2$.
For each tapping protocol, the average energy $e(F,\tau)$ of the MCSs is found to depend only on the
product $F\tau$ (see SM).

\begin{figure}[t!]
\includegraphics[width=\columnwidth]{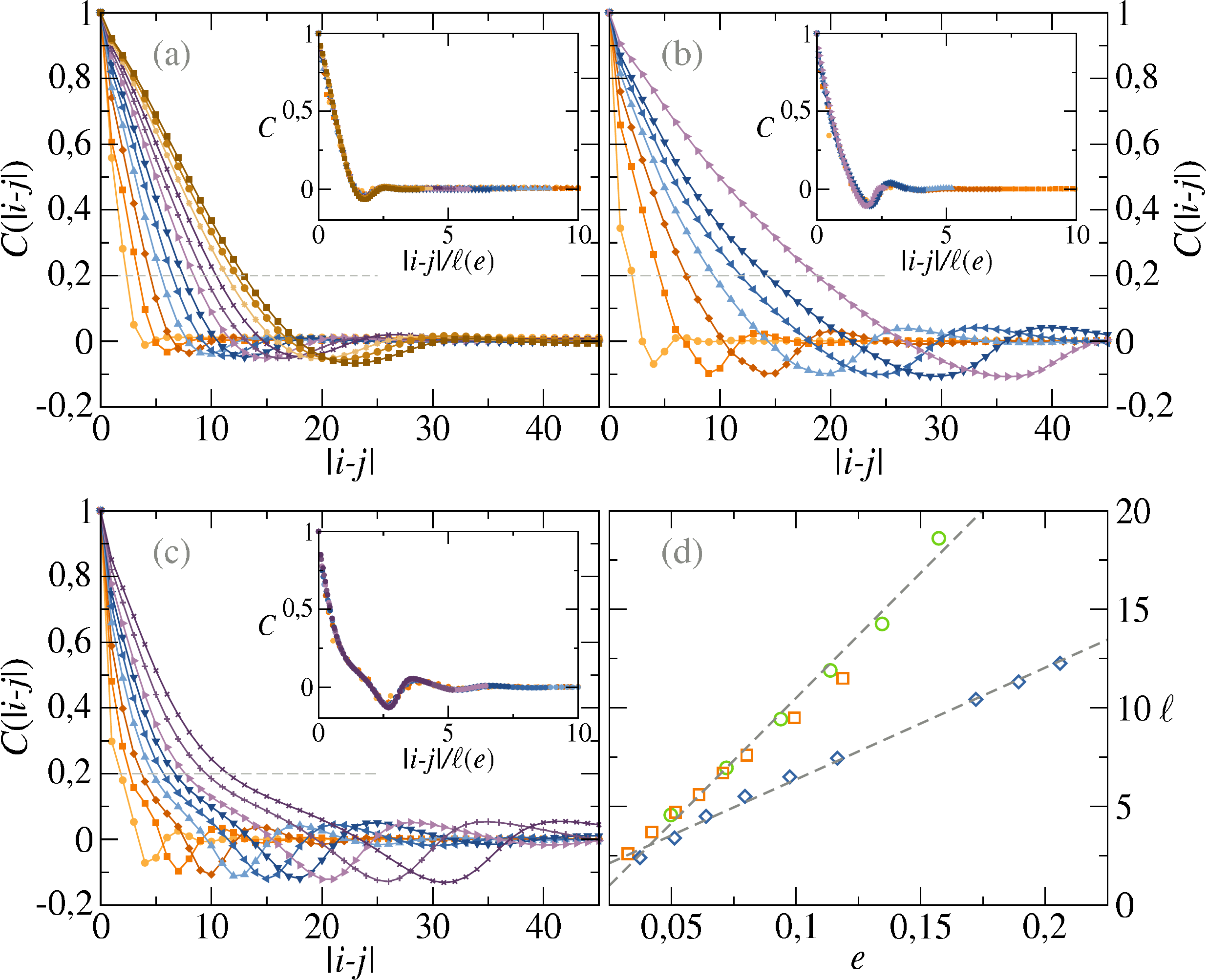}
\caption{{\bf a), b),c)}: correlation functions $C(|i-j|)$ for
  different tapping protocols, while in each panel the different
  curves correspond to different energie values. {\bf a)}
  $F\in[20,128]$, $\rho=0.3$, $\sigma=0$; {\bf b)} $F\in[20,140]$,
  $\rho=0.8$, $\sigma=0$; {\bf c)} $F\in[20,128]$, $\rho=1$,
  $\sigma=F/4$.  $\tau=60$ in all simulations.  Inset: $C(|i-j|)$
  vs.~$|i-j|/\ell(e)$, showing a good collapse of the different
  curves.  {\bf d)}: correlation length $\ell(e)$ as a function of the
  energy density $e$ of the MSC, for the different tapping protocols
  shown in a) diamonds, b) circles and c) squares, showing a linear
  increase $\ell(e)\propto e$ with a protocol-dependent slope.}
\label{fig1}
\end{figure}

To characterize the MSC we focus on the correlation function
$C(|i-j|)=\langle\xi_i\xi_j\rangle$ between the elongations of the
springs at position $i$ and $j$ in the chain.  Since this function is
trivial ($C(|i-j|=C_0\delta_{ij}$) in a thermal harmonic chain at all
temperatures, any appearance of correlations is a signature of the
unusual statistics associated with the non-Hamiltonian constraints.
Correlation functions $C(|i-j|)$ measured for different tapping
protocols are shown in Fig.~\ref{fig1}.  For a given tapping protocol
we find that the extent of correlations \emph{increases} when the
average energy of the MSC increases.  We extract the correlation
length $\ell(e)$ for each case as the distance (i.e., number of
springs) $|i-j|$ at which the measured correlation function decays
below a conventional threshold $C^*=0.2$.  The insets of
Fig.~\ref{fig1}a),~\ref{fig1}b) and~\ref{fig1}c) show the collapse of
the correlation function when the $x$ axis is rescaled with $\ell(e)$.
Fig.~\ref{fig1}d) shows, for all protocols studied, the correlation
length growing linearly with the energy $\ell(e) \sim e $; the higher
the energy, the more the system is correlated.  This result may look
puzzling at first sight, since commonly the larger is the energy the
smaller is the extent of correlations.  The key point to understand
the physics of our system is the non-Hamiltonian nature of the
constraints defining the MSC.  If for a certain $i$ we have $\xi_i\gg
1$ (a situation that is typical of a high energy MCS), then, to
fulfill the frictional constraint $|\xi_{i+1}-\xi_i|<\ms$, $\xi_{i+1}$
must be close to $\xi_i$.  The same argument relates $\xi_{i+2}$ to
$\xi_{i+1}$, and so on.  Therefore, correlations between spring
extensions build up in the MSC.  We also compare the correlation
functions characterizing the blocked states and those at the end of
the driving phase (when the force is switched off) and show that the
spring-spring correlations are not due to the external driving, but
come entirely from the constraint imposed by static friction (see SM).

\begin{figure}[t!]
\includegraphics[width=\columnwidth]{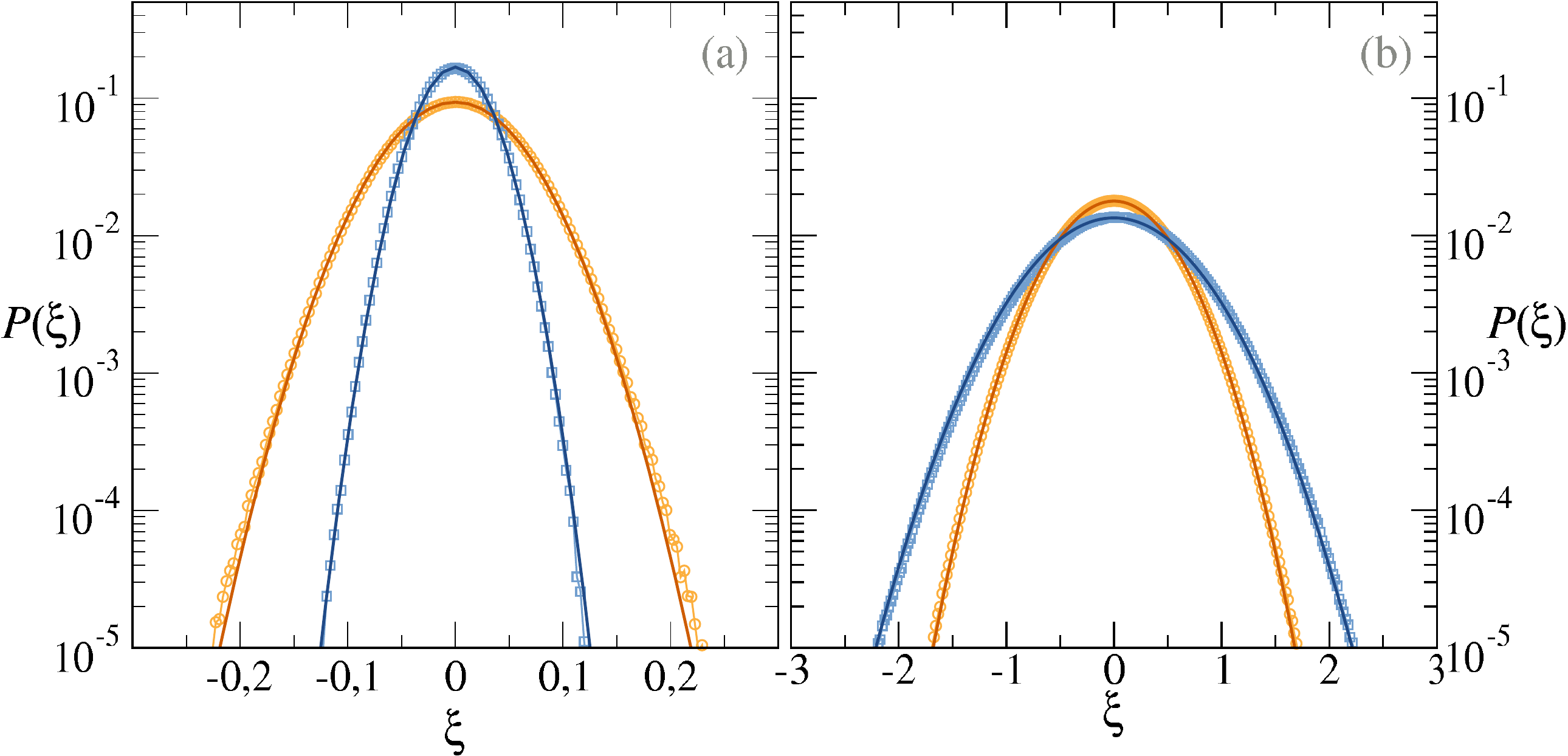}
\caption{Distributions of springs elongations $P(\xi)$ at different
  temperatures, on a semi-log scale. Points represent $P(\xi)$ in MSC
  sampled via tapping, full lines represent $P(\xi)$ obtained from Edwards theory,
  computed at the temperature $\Ted$ yielding the same energy density $e$ as the tapping.
  {\bf a)}: uncorrelated regime ($\ell(e)<1$) where $e \sim \Ted$;
  $\Ted=0.0008$ (blue squares) and $0.0026$ (orange circles).
  {\bf b)}: correlated regime ($\ell(e)>1$) where $e \sim \sqrt{\Ted}$;
  $\Ted=1.19$ (orange circles) and $3.84$ (blue squares).}
\label{fig2}
\end{figure}
It is interesting to notice that the distribution of spring lengths is Gaussian at all energies. 
This is shown in Fig.~\ref{fig2}, both in the case $\ell(e)<1$, namely when there is no correlation
between springs elongations, and in the case $\ell(e)>1$.

\paragraph{Effective theory: transfer operators--}
Given that MSC are defined in our model by the constraint $|\xi_{i+1}-\xi_i|<\ms$ $\forall i$,
from Eq.~(\ref{eq:edmes}) the probability of a configuration $\boldsymbol \xi=(\xi_1,\ldots,\xi_N)$ reads 
\be
P(\boldsymbol \xi) = e^{-\Bed\sum_{i=1}^N\xi_i^2/2}\prod_{i=1}^N \Theta(\ms-|\xi_{i+1}-\xi_i|).
\label{eq:probability_theta}
\ee
All the properties of the system can be obtained from the partition sum
$\mZ=\int_{-\infty}^{\infty} d\xi_1\ldots d\xi_N ~P(\boldsymbol \xi)$.
Using the change of variables $\xi_i=\ms\xi_i'$,
the partition function depends only (up to an irrelevant prefactor)
on the product $\Bed \ms^2$;
hence, all thermodynamic quantities are functions of $\Ted/\ms^2$.
We consider periodic boundary conditions for the chain, without imposing
any constraint on its total length.
For convenience, we fix the rest length to $l_0=\infty$, allowing us to take as domain
of integration $\xi_i\in (-\infty,\infty)$, while avoiding crossings of masses. 

Using Eq.~(\ref{eq:probability_theta}),
we have $\mZ=\Tr(\mT^N)$, with $\mT$ an operator defined as
$\mT[f](x)=\int_{-\infty}^\infty dy\, T(x,y) f(x)$, being $T(x,y)$ the
symmetric function:
\be
T(x,y)=e^{-\Bed x^2/2} \, \Theta(\ms-|x-y|)\, e^{-\Bed y^2/2}.
\label{eq:transf-op}
\ee
The operator $\mT$ has a maximum positive eigenvalue
$\lM(\Bed,\ms)$, which can be computed numerically discretizing the
domain of $\xi$, and using a complete orthonormal basis in $L^2$.
All relevant thermodynamic observables are computed in the same way (see SM).
The free-energy is obtained as $f=\Bed^{-1}\ln(\lM(\Bed,\ms))$ while the energy reads
$e=\partial(\Bed f)/\partial\Bed=-\langle\lambda_\textrm{max}|\partial\mT/\partial\Bed|\lambda_\textrm{max}\rangle/\lambda_\textrm{max}$.
In the following, we compare results from theory and simulations
by tuning the temperature $\Ted$ such that the energy $e$ takes the same value as in the numerics.

\begin{figure}[t!]
\includegraphics[width=\columnwidth]{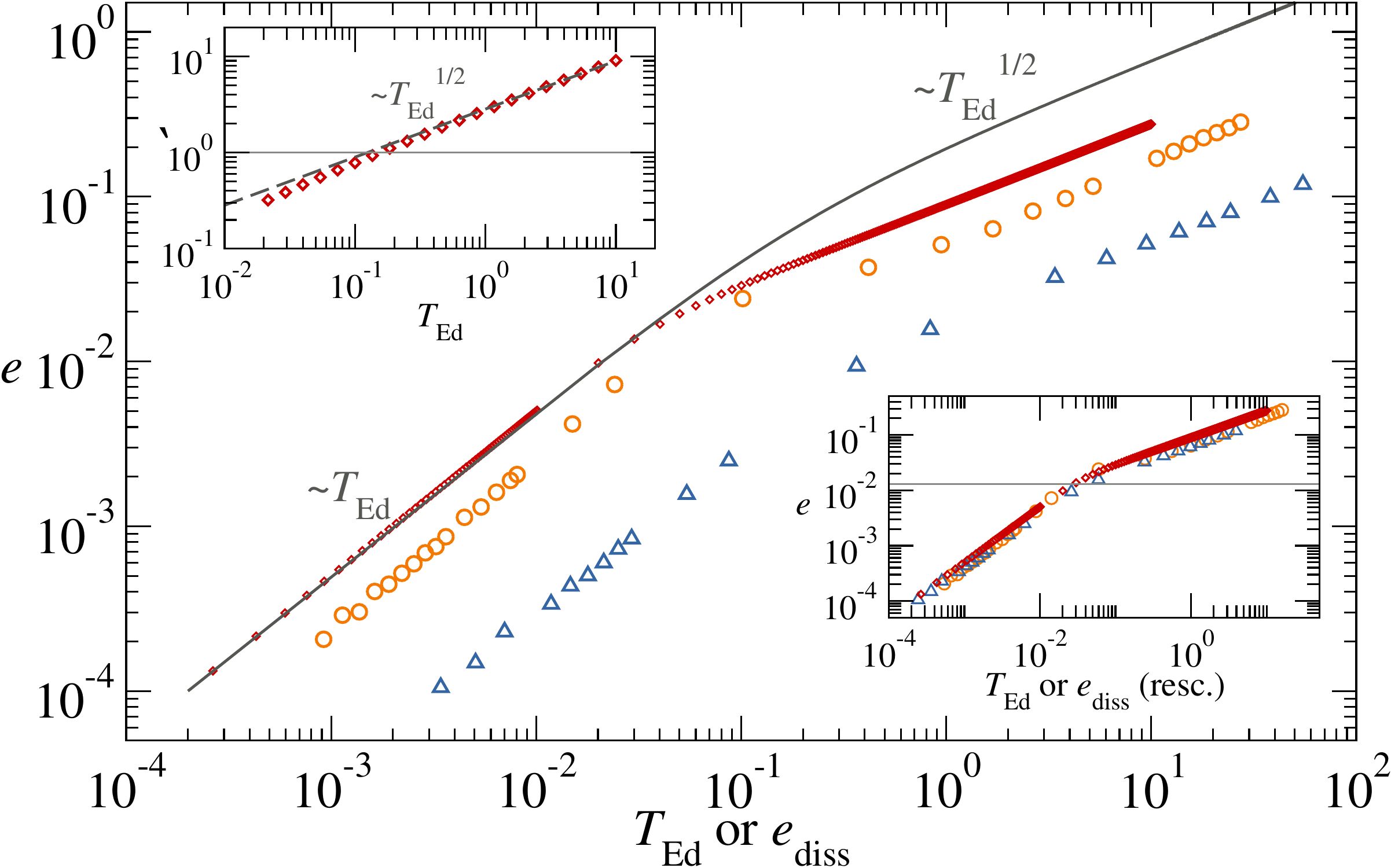}
\caption{ {\bf Main }: Energy density $e$ of MSC as a function of $\Ted$, from
  transfer operators (small red diamonds) or Gaussian approximation (full line),
  and as function of dissipated energy $e_{\textrm{diss}}$ for two tapping protocols: 1) $\rho=0.3$, $\sigma=0$ (orange circles); 2)
  $\rho=1$, $\sigma>0$ (blue triangles). For $\Ted \ll \mu^2$ one finds an ``equilibrium-like'' regime, $e \sim \Ted$;
  for $\Ted \gg \mu^2$ the behavior is $e \sim \sqrt{\Ted}$.
  {\bf Top inset}: Correlation length from exact calculation (transfer
  operators): the behavior $\ell(\Ted) \sim \sqrt{\Ted} \sim e $ is
  clear when $\Ted \gg \mu^2$.
  {\bf Bottom inset}: same symbols data sets of main panel are collapsed by just
  rescaling the $x$-axis: up to a protocol-dependent prefactor
  we have $e_{\textrm{diss}}\sim\Ted$.}
\label{fig3}
\end{figure}
The behavior of energy as a function of $\Ted$ from the transfer operator
approach is shown in Fig.~\ref{fig3}.
We find two regimes separated by a crossover that depends on $\ms$:
for $\Ted\ll \ms^2$ there is an ``equilibrium-like'' regime
where $e\sim\Ted$ while for $\Ted\gg \ms^2$ one finds $e\sim\sqrt{\Ted}$.
The transfer operator approach allows us to compute also the
probability distribution $p(\xi)$ of the elongation of a single spring (see SM).
The theoretical result for $p(\xi)$ is compared in Fig.~\ref{fig2} with the one estimated numerically from the MCSs, showing good agreement.
We find that $p(\xi)$ is
Gaussian in all regimes, even when correlations are present.

We also compute the correlation
$C(|i-j|)=\langle\xi_i\xi_j\rangle$, which is very close to an exponential form
for all values of $\Ted$, $C(|i-j|)\propto e^{-|i-j|/\ell(\Bed,\mu)}$ (see SM).
When $\Ted\gg\ms^2$ both the
correlation length $\ell$ and the energy $e$ grow as $\sqrt{\Ted}$
(see inset of Fig.~\ref{fig3}), implying $\ell\sim e$. 
We thus recover from Edwards thermodynamics the scaling behavior of correlation length
with energy observed in the simulated tapping dynamics.
This is a remarkable success of the Edwards approach for this system.
Conversely, there is almost no correlation between neighboring springs ($\ell < 1$), in the ``equilibrium-like'' low energy regime ($e\sim\Ted$).

We further show in the second inset of Fig.~\ref{fig3} that a direct measure of $\Ted$ (within a protocol dependent factor) is
obtained from the dissipated energy per tapping cycle and particle,
$e_{\mathrm{diss}}=\mu_d\langle\int_0^\tau\mathrm{sgn}(\dot{x}_i(t))\dot{x}_i(t)dt\rangle$.
Indeed, in the simulations $e_{\mathrm{diss}}$ is found to have the same scaling with $e$ as the temperature obtained
within the transfer operator approach (i.e., $e\sim e_{\mathrm{diss}}$ if $e\ll\ms^2$, $e\sim\sqrt{e_{\mathrm{diss}}}$ if $e\gg\ms^2$).
The dissipated energy can therefore be interpreted as the analog of the thermal energy that allows the system
to sample the configuration space \cite{Langer}.

\paragraph{Effective theory: Gaussian ansatz--}
To obtain approximate analytical expressions for the thermodynamic quantities, we replace the Heaviside function
in Eq.~(\ref{eq:probability_theta}) by a Gaussian function,
\be
\Theta(\ms-|\xi_{i+1}-\xi_i |) \; \rightarrow \;
\frac{1}{\sqrt{\pi}} \, \exp\left(-\frac{|\xi_{i+1}-\xi_i |^2}{4 \ms^2}\right),
\ee
yielding $\mZ \propto \int \mD{\boldsymbol\xi}
\, e^{-S(\boldsymbol \xi)}$ with an effective Hamiltonian
\be
S(\boldsymbol \xi)
= \frac{1}{2}\left[\Bed \sum_{i=1}^{N} \xi_i^2 + \frac{1}{2\ms^2}
\sum_{i=1}^N (\xi_{i+1}-\xi_i)^2\right]
\label{eq:eff-act}
\ee
This effective Hamiltonian corresponds to a positive definite quadratic form
$S(\boldsymbol\xi)=\boldsymbol\xi^T A\boldsymbol\xi/2$, where $A$ is a symmetric real
Toeplitz matrix (see SM).
The matrix A can be exactly diagonalized, yielding analytical expressions for energy, entropy,
correlation function and correlation length.
The mean energy per particle reads
\be
e(\Ted,\ms)
= \frac{1}{2} \frac{\ms \Ted}{\sqrt{2 \Ted+\ms^2}},
\label{eq:energy}
\ee
from which we recover that the crossover point between the behaviors $e\sim\Ted$ and
$e\sim\sqrt{\Ted}$ is $\Ted^*\approx\ms^2$.
In Fig.~\ref{fig3}, Eq.~(\ref{eq:energy}) is compared with the result from the transfer
operator Eq.~(\ref{eq:transf-op}), showing a semi-quantitative agreement. 
We also find that the correlation function is $\langle
\xi_i\xi_j\rangle\sim e^{-|i-j|/\ell(\Ted,\ms)}$ with
$\ell(\Ted,\ms)$ such that $\ell\sim\sqrt{T}/\ms$ in the limit $\Ted\gg\ms^2$.
This result can be recovered from a field-theoretic viewpoint, by taking a continuous
limit in Eq.~(\ref{eq:eff-act}), yielding
\be
S[\xi]
\propto \int dx \left[ \frac{1}{2} \left(\frac{\partial \xi}{\partial x}\right)^2 + \frac{1}{2} m^2 \xi^2(x) \right]
\label{eq:cont-eff-act}
\ee with a ``mass'' term $m^2=2\mu_s^2\Bed$.  The correlation function
of such a Gaussian field theory is known~\cite{LeBellac} to be
$\langle\xi(x)\xi(y)\rangle\sim e^{-m|x-y|}$, so that we recover a
correlation length $\ell \sim \Bed^{-1/2}$.  This field-theoretic
formulation confirms the presence of an infinite temperature critical
point, since the mass term goes to zero at infinite temperature.  To
inspect the critical exponents associated with this critical point, we
study the fluctuations of the total energy of the chain, $\delta
E=\frac{1}{2}\sum_{i=1}^N(\xi_i^2-\langle \xi_i^2\rangle)$ and the
fluctuations of its total length $\delta \mL =\sum_i \xi_i$.  We find
(see SM) that the variance of both energy and length diverge linearly
with temperature (or, equivalently, as $\ell^2$),
\be \label{eq:crit-exp}
\frac{\langle [\delta E]^2 \rangle}{N} \sim
\frac{\langle [\delta \mL]^2 \rangle}{N} \sim \Ted \,.
\ee
Finally, we
compute the entropy density $s=-\partial f/\partial\Ted$.  We find
that it saturates at high temperature to a finite value,
$\lim_{\Ted\rightarrow \infty} s(\Ted,\ms) = \frac{1}{2}\ln
2+\ln\mu_s$.  This saturation results from the presence of long-range
correlation at infinite temperature.  This can be confirmed by
contrast, computing the ``mean-field'' entropy density $s_{\rm
  mf}=-\int d\xi\,p(\xi)\,\ln p(\xi)$, with $p(\xi)$ the distribution
of a single spring elongation $\xi$.  We find that $s_{\rm mf}$, which
discards correlations, diverges like $\ln \Ted$ at infinite
temperature (see SM), at odds with the saturation of the entropy $s$.

\paragraph{Conclusions--} 
The present study provides a clearcut example of how an effective thermodynamic theory can successfully
describe an \emph{athermal} dissipative system.  
The most remarkable difference between standard equilibrium thermodynamics and the effective theory we have presented
is the presence of an infinite temperature critical point, with an associated divergence of the correlation length as
$\ell \sim \Ted^{1/2}$ (or $\ell \sim e$).
As seen in the field-theoretic formulation Eq.~(\ref{eq:cont-eff-act}), this infinite temperature critical point results
from the long-range correlation generated by static friction in the blocked states.
The difference with standard equilibrium systems is that the gradient term in the effective Hamiltonian does not come
from an energetic interaction, but from a non-Hamiltonian constraint.
Its coefficient is strictly temperature independent,
while the coefficient of the energetic term scales inversely with temperature.
While temperature-independent terms could also be present at equilibrium (e.g., entropic constraints such as excluded volume),
they are usually purely local and do not involve gradient terms.
Hence, in spite of its simplicity, our model exhibits a phenomenology clearly distinct from that of equilibrium systems,
and the field-theoretic formulation suggests that the results should be quite robust to changes in the details of the model.
Future work should investigate this issue in more details.

\begin{acknowledgments}
We acknowledge Financial support from ERC Grant No. ADG20110209. GG
thanks A. Cavagna, A. Puglisi and A. Vulpiani for useful discussions
and comments.
\end{acknowledgments}

\newpage 
\appendix
\setcounter{equation}{0}

\onecolumngrid

\section*{SUPPLEMENTARY MATERIAL:\\
E\MakeLowercase{dwards thermodynamics for a driven athermal system with dry friction}}

\twocolumngrid
\subsection{Numerical simulations}
\subsubsection*{Details on the simulations}
The tapping protocol is defined in Eq.~(3) in the main text.
The motivation for introducing (annealed) disorder in the tapping protocol is that, when taking $\rho=1$ and $\sigma=0$ (i.e., without any disorder in the driving), we found undamped waves
which travel across the chain causing an undesired and artificial
collapse of the ``ensemble'' of visited MSC to a very small and specific subset.
As mentioned in the main text, for the numerical simulations of the
tapping dynamics we consider \emph{open} boundary conditions. This
choice is done in order to avoid constraints on springs which may
induce trivial correlations.  For instance if one fixes
$\sum_{i=1}^{N}\xi_i = N l_0 $, then also has $N^2 l_0^2 = \langle L^2
\rangle = \sum_{ij} \langle \xi_i \xi_j\rangle$, which is a constraint
on the two point correlation function.
With open boundary conditions the first and the
last blocks are connected to a single spring and their equations of
motions are respectively 
\bea
\ddot{x}_1 &=&-\md \textrm{sgn}{\dot{x}_1} + (x_2-x_1-l_0) + f^{\textrm{ext}}_1 \nonumber \\ 
\ddot{x}_{N+1} &=& -\md \textrm{sgn}{\dot{x}_{N+1}} - (x_{N+1}-x_N-l_0) + f^{\textrm{ext}}_{N+1} 
\eea 
Note that in order to have $N$ springs with open
boundary conditions we need $N+1$ blocks.

\subsubsection*{Dependence of the energy on driving parameters}

\begin{figure}
\includegraphics[width=\columnwidth]{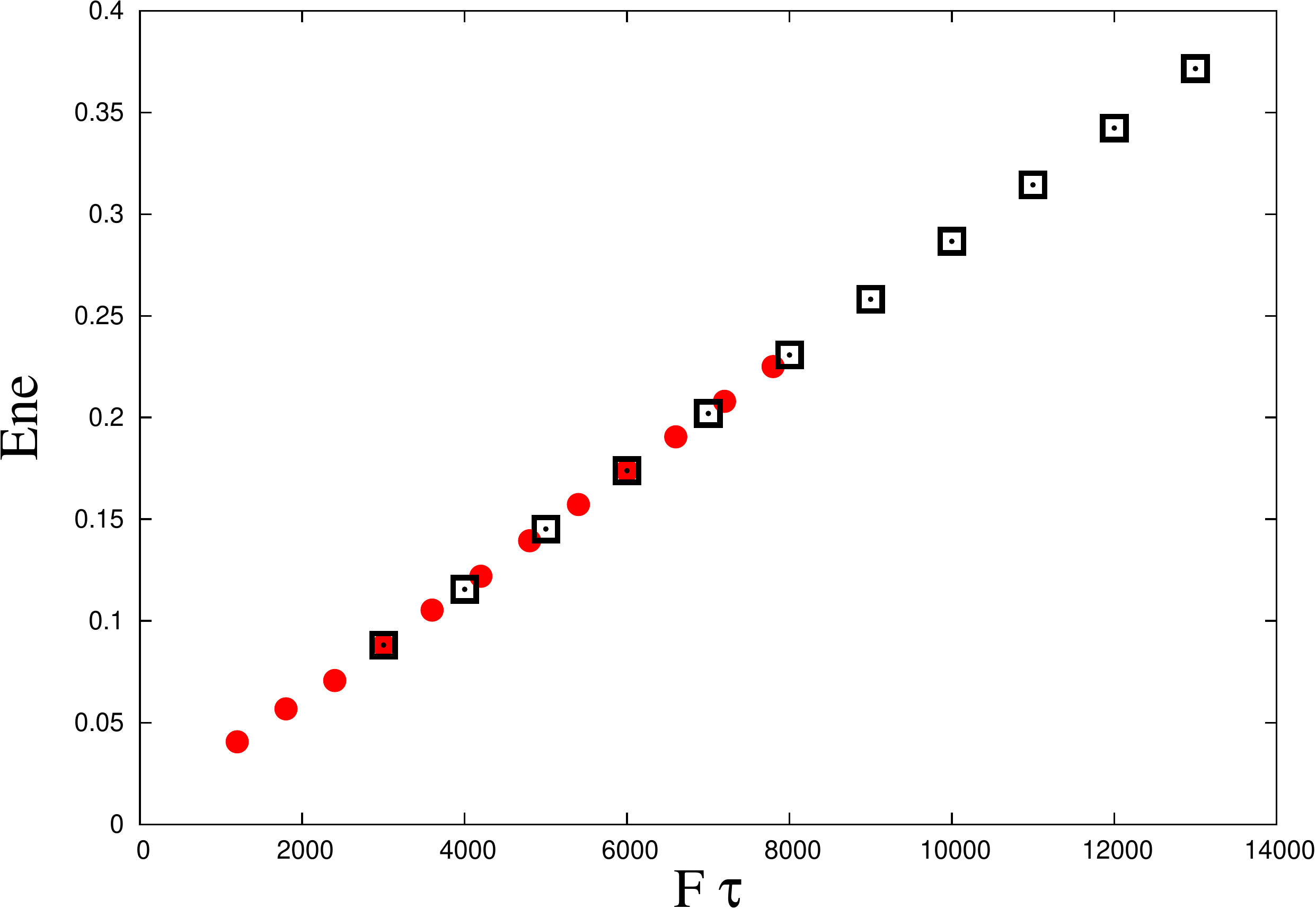}
\caption{Energy per spring stored in mechanically stable
  configurations (MSC) as function of $F\tau$ with $F$ the pulling
  force and $\tau$ the duration of the pulling. Full circles: $F$ is
  increased at fixed $\tau$; Empty squares: $\tau$ is increased at
  fixed $F$.}
\label{fig0}
\end{figure}

The average energy of MSC depends in
principle on the four parameters which charaterize the driving cycle: 
the duration $\tau$ of forcing, the intensity of the mean applied force $F$,
its standard deviation $\sigma$ as well as the fraction $\rho$ of pulled particles. The parameters $\rho$ and $\sigma$ define the tapping protocol, while $F$ and $\tau$ characterize the intensity of the driving.
We have found that for a fixed tapping protocol,
the energy $e$ of the dynamically reached MSC does not depends separately on $F$ and $\tau$, but is uniquely determined by the product $F\tau$. This is
clearly shown by the set of data reported in Fig.~\ref{fig0}.

\subsection{One-cycle experiment: test for the origin of friction induced correlations}
In the manuscript we presented data on the ``spring-spring''
correlation function $\langle \xi_i \xi_j \rangle$ obtained in the
mechanically stable configurations (MSCs). Due to the non-random type
of the external force $F$ used to sample different MSCs the reader is
allowed to wonder if the correlations we measure in the MSCs are
really induced by friction at $100\%$, or a part of them is induced by
the same $F$. The numerical evidence that it is not the case comes
from the ``one-cycle driving experiment'' we are going to discuss. The
one-cycle driving experiment is analogous to the tapping dynamics
discussed in the main text apart from one thing: at the beginning of
each driving cycle the positions of blocks are randomized.  More
precisely all the $x_i$ are reset to $x_i = i~l_0 + dx$, where $dx$ is
normally distributed and $l_0$ is the rest length of springs. During
the elementary driving cycle of the dynamics a fraction $\rho$ of the
particles is pulled for a durantion $\tau$ with a constant force $F$,
the force is then switched off and the system relaxes to a MSC. We
average over many iterations two correlation functions: a) the
\emph{forced} correlation $\langle \xi_i \xi_j \rangle_F$, which is
measured just before switching-off the force at $\tau$; b) The
correlation in the mechanically stable configurations reached after
the quench, $\langle \xi_i \xi_j \rangle_{\textrm{MSC}}$. Our aim in
studying $\langle \xi_i \xi_j \rangle_F$ is to see \emph{how much} of
the correlation is provided by the external force. On the other hand
we must also be sure that when measuring $\langle \xi_i \xi_j
\rangle_F$ there is no memory of the correlation in $\langle \xi_i
\xi_j \rangle_{\textrm{MSC}}$ at the previous step. It is for this
reason that at the beginning of each driving cycle the positions of
block are randomized. In this way we make ourselves sure that if some
correlation is measured in $\langle \xi_i \xi_j \rangle_F$, it comes
solely from the external force. The result is illustrated in
Fig.~\ref{fig1}. The continuous line is $\langle \xi_i \xi_j
\rangle_{\textrm{MSC}}$ while the (red) circles represent $\langle
\xi_i \xi_j \rangle_F$: being the latter identically zero we can
conclude that no correlation is induced by the force and that the
correlations we observe in the system come solely from the constraint
imposed by static friction on the MSCs.
\begin{figure}
\includegraphics[width=\columnwidth]{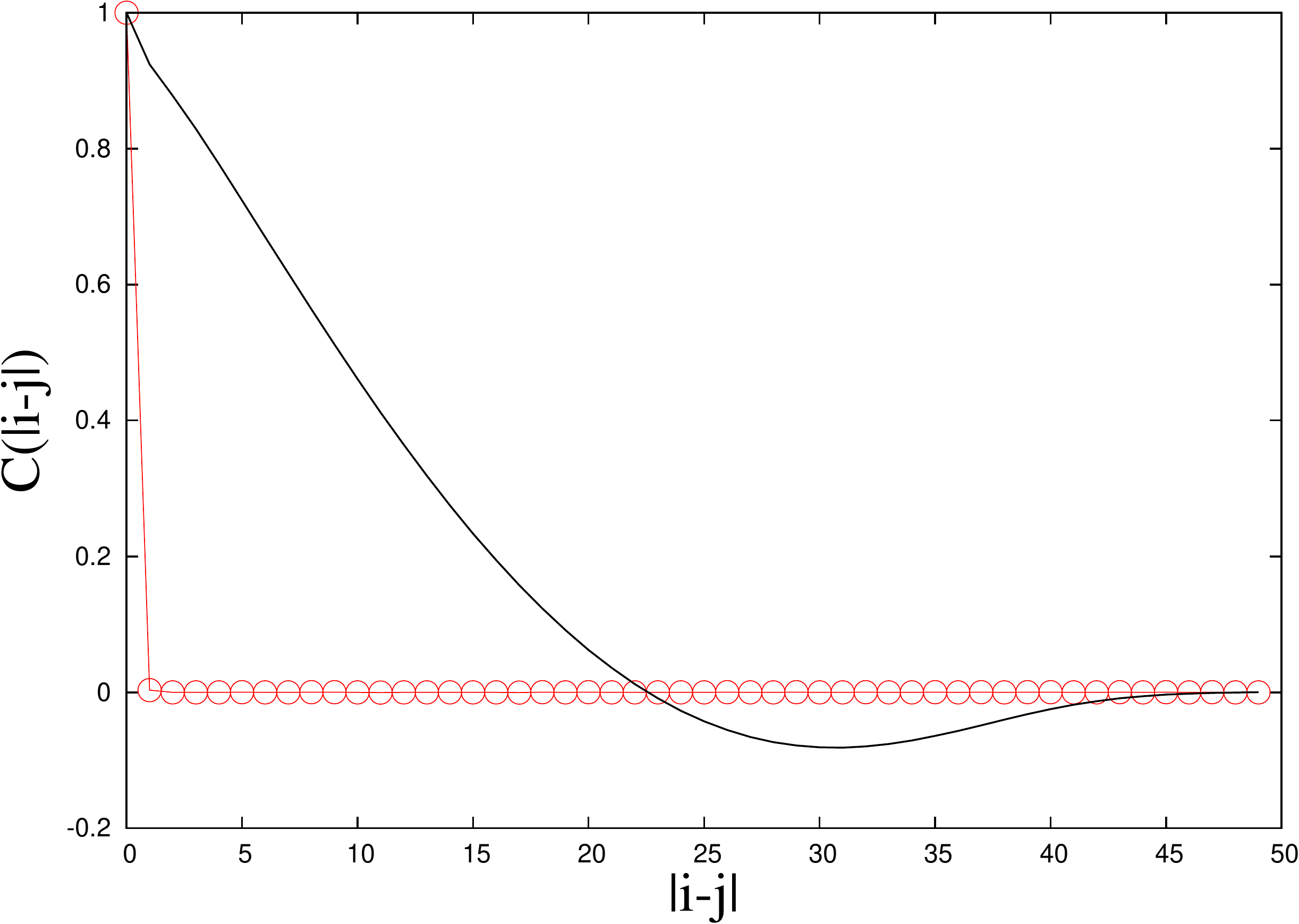}
\caption{Spring-spring correlation function $\langle \xi_i \xi_j
  \rangle_F$ (red circles) in presence of the external force and
  $\langle \xi_i \xi_j \rangle_{\textrm{MSC}}$ (continuous line)
  measured in the mechanically stable configurations reached after
  quench. Data are obtained within a ``one-cycle-driving-experiment'':
  an external constant force $F=100$ is applied for a duration
  $\tau=100$ pulling a fraction $\rho=0.25$ of the blocks. The number
  of blocks is $N=256$. At the beginning of each driving cycle the
  positions of the blocks are randomized.}
\label{fig1}
\end{figure}
\begin{figure}[h!]
\includegraphics[width=\columnwidth]{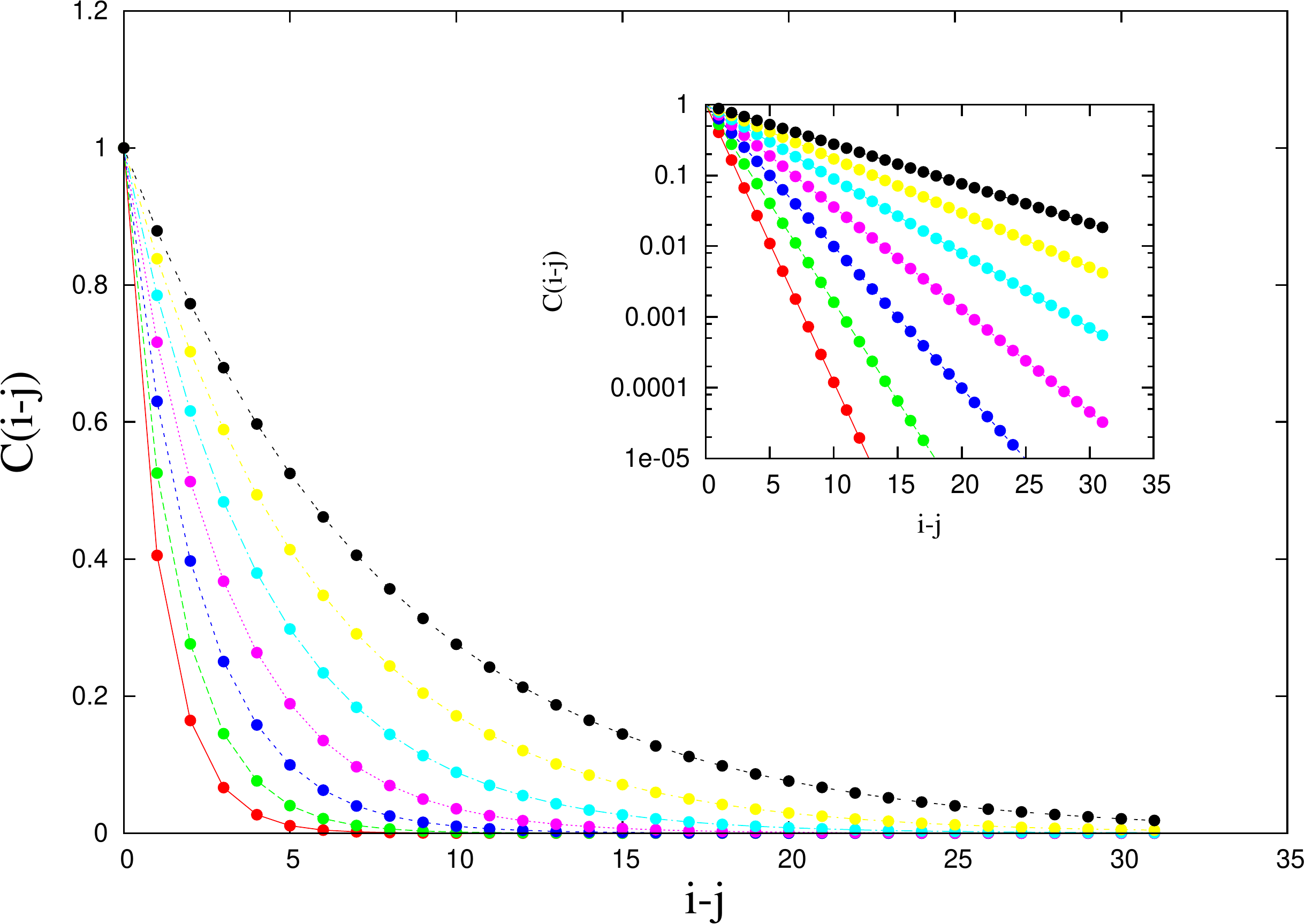}
\caption{{\bf Main}: spring-spring correlation function $C(i-j)$, where $i$
  and $j$ labels different springs ($i>j$), computed according to
  Eq.~(\ref{eq:correlation}) at different temperatures. From left to
  right: $T=0.185,0.341,0.631,1.166,2.154,3.981,7.356$. At odds with
  what usually happens the extent of correlations increases as the
  temperature is increased. {\bf Inset}: $C(i-j)$ vs $i-j$ in semi-log
  scale: all correlations decay exponentially.}
\label{fig00}
\end{figure}
\subsection{Transfer operators}
The core of the exact solution of the effective thermodynamics of our
model is represented by the spectral properties of the linear operator
$\mT$, acting in the space of square integrable functions $L^2$,
$\mT:L^2\rightarrow L^2$. The operator $\mT$ is defined as follows:
\be \mT[f](x) = \int_{-\infty}^{\infty} dy~ T(x,y) f(y), \ee
with \be T(x,y)=e^{-\frac{\Bed}{4}y^2} \Theta(\ms-|y-x|)
e^{-\frac{\Bed}{4}x^2}.  \ee As can be easily checked $\int dx dy
|T(x,y)|^2 < \infty$, so that $\mT$ is a Hilbert-Schmidt operator, and
hence compact and bounded, which guarantees the existence of a maximum
positive eigenvalue. The kernel is symmetric and real,
$T(x,y)=T^*(y,x)$, so that the operator is also self-adjoint, which
guarantees the spectral theorem to hold: a complete set of eigenvalues
and eigenvectors exist for our operator~\cite{fonseca}.
\subsubsection*{Correlation function}
By means of the transfer operator approach we can compute several
quantities of interest: the one that is most relevant for our study is
the correlation between the elongation of far apart springs.  Let us
consider for instance the elongations $\xi_i$ and $\xi_j$, with $j > i
$. In terms of our effective theory the correlation between these two
elongations can be written as: \bea \langle \xi_i \xi_j \rangle &=&
\frac{1}{\mathcal{Z}} \int d\xi_1 \ldots \xi_N T(\xi_1,\xi_2)\ldots
T(\xi_{i-1},\xi_i) \xi_i T(\xi_i,\xi_{i+1})\nonumber \\ &\dots&
T(\xi_{j-1},\xi_j) \xi_j T(\xi_j,\xi_{j+1})\ldots T(\xi_N,\xi_1), \nonumber \\
\label{eq:correlation00}
\eea
which in terms of operators becomes 
\be \langle \xi_i \xi_j \rangle =
\frac{1}{\mathrm{Tr}[\mT^N]} \mathrm{Tr}\left[ \mT^{i-1} \mD \mT^{j-i}
  \mD \mT^{N-j+1} \right],
\label{eq:correlation0}
\ee
where $\mD$ is the diagonal operator defined as:
\be
\mathcal{D}[f](x)= x f(x).
\label{eq:diagonal}
\ee
Exploiting the fact that the operator $\mT$ has a complete
spectrum of orthonormal eigenvectors we can use the completeness
relation $\boldsymbol 1 = \sum_a |a \rangle \langle a|$, where $|a
\rangle$ is the eigenvector of $\mT$ relative to the eigenvalue $a$,
to simplify Eq.~(\ref{eq:correlation0}). One can write the trace in
Eq.~(\ref{eq:correlation0}) as $\mathrm{Tr}(\mT) = \sum_a \langle a|
\mT|a \rangle$ and then insert between each pair of neighboring
operators the completeness $\boldsymbol 1 = \sum_a |a \rangle \langle
a|$. The result is
\begin{eqnarray}
&& \langle \xi_i \xi_j \rangle = \nonumber \\ && \frac{1}{\sum_a
  \lambda_a^N} \sum_{a,b,c,d,e} \langle a| T^{i-1} | b \rangle \langle
b| D | c \rangle \langle c| T^{j-i} | d \rangle \nonumber \\ && \langle d| D | e
\rangle \langle e| T^{N-j+1} |a\rangle \nonumber \\ &=&
\frac{1}{\sum_a \lambda_a^N} \sum_{a,b} \lambda_a^{i-1} \langle a| D |
b \rangle \lambda_b^{j-i} \langle b| D | a \rangle \lambda_a^{N-j+1},
\end{eqnarray}
which in the thermodynamic limit becomes
\begin{eqnarray}
\lim_{N\rightarrow\infty} \langle \xi_i \xi_j \rangle &=& \frac{1}{\lM^N} \sum_{b} \lM^{N-j+i} \langle \lM | D | b \rangle \lambda_b^{j-i} \langle b| D | \lM \rangle  \nonumber \\
&=& \sum_{b} \left(\frac{\lambda_b}{\lM}\right)^{j-i} |\langle b| D |\lM \rangle|^2  \nonumber \\
&=& \sum_{b} \left(\frac{\lambda_b}{\lM}\right)^{j-i} \left|\int_{-\infty}^{\infty} dx ~x f_{ _b}(x) f_{ _{\lambda_\textrm{max}}}(x) \right|^2  \nonumber, \\
\label{eq:correlation}
\end{eqnarray}
where $f_{ _b}(x)$ is the eigenfunction related to the eigenvalue $b$.
The correlation functions obtained at different values of the Edwards temperature
$\Ted$ by means of a numerical determination of eigenvalues and eigenvectors of $\mT$ 
can be found in Fig.~\ref{fig00}: at all temperatures, we numerically observe an exponential decay, meaning that the second largest eigenvalue dominates the sum in Eq.~(\ref{eq:correlation}).
\subsubsection*{Energy}
The average energy of the harmonic chain with $N$ particles is defined
as:
\begin{eqnarray}
e_N(\Bed) &=& \frac{\partial}{\partial\Bed}\left( \Bed f\right) = -\frac{1}{N} \frac{\partial}{\partial\Bed} \ln(\mathcal{Z}) \nonumber \\
&=& -\frac{1}{N~\Tr[\mT^N(\Bed)]}~\frac{\partial}{\partial\Bed} \Tr[\mT^N(\Bed)] \nonumber \\
&=& -\frac{1}{\Tr[\mT^N(\Bed)]} \Tr[\mT^{N-1}(\Bed)\frac{\partial \mT}{\partial\Bed}],
\label{eq:energy}
\end{eqnarray}
where in the last line of Eq.~(\ref{eq:energy}) we passed the derivative
sign $\partial/\partial\Bed$ under the trace sign. The ``matrix elements'' of the operator $\partial
\mT/\partial \Bed$ are simply defined as the derivative with respect
to $\Bed$ of the matrix elements of $\mT$:
\be
\frac{\partial T}{\partial\Bed}(x,y) = - \frac{x^2 + y^2}{4} \exp\left( - \frac{\Bed}{4} [x^2 + y^2] \right).
\ee
Exploiting again the definition of the trace $\mathrm{Tr}[\mT^N] =
\sum_a \langle a | \mT^N | a \rangle$ and by inserting the
completeness relation ${\bf 1}=\sum_a |a\rangle \langle a |$, 
we can write
\begin{eqnarray}
e_N(\Bed)&=& -\frac{1}{\Tr[\mT^N(\Bed)]} \Tr[\mT^{N-1}(\Bed)\frac{\partial \mT}{\partial\Bed}] \nonumber \\ 
&=& -\frac{1}{\Tr[\mT^N(\Bed)]} \Tr[\mT^{N-1}(\Bed) \cdot {\bf 1} \cdot \frac{\partial \mT}{\partial\Bed}] \nonumber \\
&=&-\frac{1}{\sum_a \lambda_a^N} \sum_{ab} \langle a | \mT^{N-1}| b \rangle \langle b | \frac{\partial \mT}{\partial\Bed}| a \rangle \nonumber \\ 
&=& -\frac{1}{\sum_a \lambda_a^N} \sum_a \lambda_a^{N-1} \langle a | \frac{\partial \mT}{\partial\Bed} | a \rangle \nonumber \\
\end{eqnarray}
which in the thermodynamic limit yields 
\begin{eqnarray}
\lim_{N\rightarrow \infty} e_N(\Bed) &=& -\frac{1}{\lambda_\textrm{max}}  \langle \lambda_\textrm{max} | \frac{\partial \mT}{\partial\Bed} | \lambda_\textrm{max} \rangle \nonumber \\ 
&=& -\frac{1}{\lambda_\textrm{max}} \int_{-\infty}^{\infty} dx f_{ _{\lambda_\textrm{max}}}(x) \partial_{\Bed}\mT(x,y) f_{ _{\lambda_\textrm{max}}}(y), \nonumber \\
\end{eqnarray}
where $f_{\lambda_\textrm{max}}(y)$ is the eigenfunction related to
the maximum eigenvalue.
\subsubsection*{Marginal distributions}
By means of the transfer matrix technique is easy to obtain marginal
distributions. For instance we are interested in the probability
distribution of the springs elongation, which reads
\bea
P_N(\xi) &=&
\frac{1}{\mathcal{Z}} \int d\xi_2 \ldots \xi_N ~T(\xi,\xi_2)\ldots
T(\xi_N,\xi) \nonumber \\
&=& \frac{1}{\mZ} \langle \xi| \mT^N | \xi
\rangle \nonumber \\ 
&=& \frac{1}{\sum_a \lambda_a} \sum_{ab} \langle
\xi | a \rangle \langle a |\mT^N | b \rangle \langle b | \xi \rangle
\nonumber \\
&=& \frac{1}{\sum_a \lambda_a^N} \sum_a \lambda_a^N
|\langle a | \xi \rangle|^2 \nonumber \\
\lim_{N \rightarrow \infty} P_N(\xi) &=& |\flM(\xi)|^2,
\eea
where, again, $\flM(x)$ is the
eigenfunction related to the maximum eigenvalue. This distribution,
which has a Gaussian shape, is plotted for different values of $\Ted$
in Fig.2 in the Letter.
\subsection{Gaussian approximation}
An explicit computation of how the internal energy depends on the
temperature is possible if one assumes a Gaussian smoothening of the
Heaviside function enforcing the static friction constraint:
\be
\Theta(\ms-|\xi_{i+1}-\xi_i |) \Longrightarrow \frac{1}{\sqrt{\pi}}
\exp\left(-\frac{|\xi_{i+1}-\xi_i |^2}{4 \ms^2}\right)
\ee
With this approximation the whole partition function reads  
\bea \nonumber
\mathcal{Z}  &=&  \pi^{-N/2} \int \mathcal{D}\boldsymbol \xi e^{-S(\boldsymbol \xi)}
\\
&=& \pi^{-N/2} \int d\xi_1 \ldots d\xi_N ~~\exp\left[ -\frac{1}{2}\boldsymbol \xi \cdot \mathbf{A} \boldsymbol \xi \right]
\label{eq:GaussInt}
\eea
where $\boldsymbol \xi$ represents the vector  $\boldsymbol \xi = (\xi_1,\ldots,\xi_N)$ and $\mathbf{A}$ is a $X\times N$ symmetric real Toeplitz matrix of the kind 
\begin{equation*}
\mathbf{A} = \left(
\begin{array}{cccccc}
A_0 & A_1 & 0 & 0 & 0 & A_1 \\
A_1 & A_0 & A_1 & 0 & 0 & 0 \\
0 & A_1 & A_0 & A_1 & 0 & 0  \\
0 & 0 & A_1 & A_0 &  A_1& 0  \\
0 & 0 & 0 & A_1 & A_0 & A_1  \\
A_1 & 0 & 0 & 0 & A_1 & A_0  \\
\end{array} \right)
\end{equation*}
with 
\bea 
A_0 &=& \Bed+\frac{1}{\ms^2} \nonumber \\ 
A_1 &=& - \frac{1}{2\ms^2} 
\eea 
For this kind of matrix the eigenvalues can be explicilty written as 
\bea 
\lambda_k(\Bed,\ms) &=& A_0 + 2 A_1 \cos\left( \frac{2 \pi k }{N} \right) \nonumber \\ 
&=& \Bed+\frac{1}{\ms^2} \left[ 1 - \cos\left( \frac{2 \pi k }{N} \right) \right].  
\eea
\subsubsection*{Free energy, energy and entropy}

The free-energy reads then as
\bea 
f &=& -\frac{1}{\Bed N} \ln\left( \pi^{-N/2}\int d\xi_1 \ldots d\xi_N ~~\exp\left[ -\frac{1}{2}\boldsymbol \xi \cdot A \boldsymbol \xi \right] \right) \nonumber \\ 
&=& \frac{\ln 2}{2\Bed} + \frac{1}{2\Bed N} \ln\det(A) \nonumber \\ 
&=& \frac{\ln 2}{2\Bed}  + \frac{1}{2\Bed N} \sum_k \ln[\lambda_k(\Bed,\ms)].  
\eea 
In the continuous limit
\be
\frac{1}{N} \sum_{k=1}^{N} \sim \int_0^1 dk
\ee
the free-energy can be exactly computed and reads: 
\bea
f &=& \frac{\ln 2}{2\Bed}  + \frac{1}{2\Bed} \int_0^1 dk ~\ln\left(
\Bed+\frac{1}{\ms^2} \left[ 1 - \cos\left(2 \pi k \right)
  \right] \right) \nonumber \\
&=& \frac{\ln 2}{2\Bed} +
\frac{1}{2\Bed} \left[ \ln\left( \frac{\Bed}{4}\right) +
  2\ln\left( 1 + \sqrt{1+\frac{1}{\Bed\ms^2}}\right) \right]\nonumber \\
\label{eq:exact-free-enetemp}
\eea
Note that this method to compute the free-energy is different from the transfer operator method. Here we do not write the partition function as the trace of the $N^{\rm th}$ power of a transfer operator, but we diagonalize exactly the quadratic form appearing in Eq.~(\ref{eq:GaussInt}). The partition function is then exactly computed for all $N$ using the Gaussian integral formula, and all eigenvalues of $\mathbf{A}$ contribute to the free-energy (while only the largest eigenvalue of the transfer operator contributes to the free energy in the thermodynamic limit).

The energy per degree of freedom can also be computed:
\bea 
e = \frac{\partial (\Bed f)}{\partial \Bed} &=& \frac{1}{2 N}
\sum_{k=1}^{N} \frac{1}{\Bed + \frac{1}{\ms^2} \left[ 1 - \cos\left(
    \frac{2 \pi k }{N} \right) \right]} \nonumber \\ &\sim&
\frac{1}{2} \int_0^1 \frac{d k}{\Bed + \frac{1}{\ms^2} \left[ 1 -
    \cos(2 \pi x) \right]} \nonumber \\ &=& \frac{\ms}{2
  \sqrt{2\Bed+\Bed^2 \ms^2}} = \frac{1}{2} \frac{\ms \Ted}{\sqrt{2
    \Ted+\ms^2}}.\nonumber \\ 
\label{eq:exact-enetemp}
\eea
Combining Eqs.~(\ref{eq:exact-free-enetemp}) and (\ref{eq:exact-enetemp}),
the entropy per degree of freedom can be simply obtained as $s(\Bed) = \Bed e(\Bed) - \Bed f(\Bed)$. 
One can check that $s(\Bed) \to \frac{1}{2}\ln 2 + \ln \ms$ when $\Ted \to \infty$.
Also, energy fluctuations are given by
\be
\frac{\langle [\delta E]^2 \rangle}{N} =
-\frac{\partial^2}{\partial\beta^2}(\beta f) 
= \frac{\ms}{2} \, \frac{1+\Bed\ms^2}{2\Bed+\Bed^2\ms^2},
\ee
and at high temperature
they diverge as $\langle [\delta E]^2 \rangle/N \sim \Ted$, consistently with Eq.~(10) in the main text.

\subsubsection*{Correlation function}

When the probability of a configuration of the system is a multivariate gaussian of the kind 
\be
P(\boldsymbol \xi) \propto \exp\left( -\frac{1}{2} \sum_{ij} A_{ij} \xi_i \xi_j\right)
\ee 
we know that the value of the average correlation between $\xi_i$ and
$\xi_j$ is 
\be 
\langle \xi_i \xi_j \rangle = \left( A^{-1} \right)_{ij}.
\ee 
For the problem we are studying $A$ is a tridiagonal Toeplitz matrix, for which the 
explicit formulae for the elements of the inverse is~\cite{fonseca}: 
\be
\left( A^{-1} \right)_{ij} = (-1)^{i+j} \frac{A_1^{j-i}}{|A_1|^{j-i+1}} \frac{U_{i-1}(d)U_{n-j}(d)}{U_n(d)},
\label{eq:inverse}
\ee
where $d=A_0/(2|A_1|)$, and $U_n(x)$ is a Chebyshev polynomial of the second kind of degree $n$ in 
the variable $x$. For such polynomials, a useful identity is known:
\be
U_n(x) = \frac{\sin \theta (n+1)}{\sin \theta}, 
\ee
where $\theta$ is the complex number such that $x=\cos(\theta)$ and $0\leq\Re(\theta)<\pi$.
In our case we have 
\be
d = A_0/(2|A_1|) = 1 + \ms^2 \Bed > 1, 
\ee
where the last inequality on the right is true for every finite value
of the temperature and of the friction coefficient. Therefore the number
$\theta$ must be a purely imaginary one, i.e., $\theta = i \alpha$,
with $\alpha>0$. For our problem we have therefore that the Chebyshev
polynomials are defined according to the two following identities:
\bea
U_n(1 + \ms^2\Bed) &=& \frac{\sinh \alpha (n+1)}{\sinh \alpha} \nonumber \\ 
1 + \ms^2\Bed &=& \cosh(\alpha)
\eea
Because our problem is in the continuum we are interested in the limit $n\rightarrow \infty$ 
of the expression in Eq.~(\ref{eq:inverse}). Let us notice the following: 
\bea
\lim_{n\rightarrow\infty} \frac{U_{n-j}(d)}{U_n(d)} &=& \lim_{n\rightarrow\infty} \frac{\sinh[(n-j+1)\alpha]}{\sinh[(n+1)\alpha]} = e^{-\alpha j} \nonumber \\
U_{i-1}(d) &=& \frac{\sinh(i\alpha)}{\sinh(\alpha)} \nonumber \\
\lim_{\substack{i\rightarrow\infty \\ i<j}} \frac{U_{i-1}(d)U_{n-j}(d)}{U_n(d)} &=& \lim_{\substack{i\rightarrow\infty \\ i<j}}  \frac{e^{-\alpha(j-i)}-e^{-\alpha(i+j)}}{2\sinh(\alpha)} \nonumber \\ 
&=& \frac{e^{-(j-i)/\ell}}{2\sinh(\ell^{-1})}, 
\eea
where the correlation length $\ell$ in the last row is defined as $\alpha=\ell^{-1}$. In the limit where $\ms^2\Bed\ll 1$, namely when 
the temperature is large compared to the friction coefficient, the correlation length $\ell$ must be large and we can estimate its asymptotic scaling:
\bea 
\cosh(\ell^{-1}) &=& \ms^2\Bed +  1 \nonumber \\
\ms^2\Bed \ll 1 ~~~ &\Longrightarrow&  \ms^2\Bed +  1 \sim 1 + \frac{\ell^{-2}}{2} \nonumber \\
\ms^2\Bed \ll 1 ~~~ &\Longrightarrow& \ell \sim \frac{\sqrt{\Ted}}{\ms}
\eea 
By recalling that $A_1=-1/(2\ms^2)$, we can finally write the two-point correlation function as:
\bea
\langle \xi_i \xi_j \rangle &=& (A^{-1})_{ij} = \frac{\ms^2}{\sinh(\alpha)} e^{-\alpha(j-i)} \nonumber \\ 
&=& \frac{\ms^2\Ted}{\sqrt{\ms^4+2 \Ted\ms^2}} e^{-(j-i)/\ell(\ms,\Ted)} \nonumber \\ 
\langle \xi_i \xi_j \rangle &=& 2 ~e(\ms,\Ted)~ e^{-(j-i)/\ell(\ms,\Ted)},
\label{eq:corr-SM}
\eea
where the correlation length is 
\be
\ell(\ms,\Ted) = \frac{1}{\mathrm{arcosh}(1+\ms^2/\Ted)}.
\ee
which scales for $\Ted \to \infty$ as
$\ell(\ms,\Ted) \sim \sqrt{\Ted}/\ms$.

The variance of the fluctuations of the total length $\mathcal{L}=\sum_{i=1}^N$ read
\be
\langle [\delta \mathcal{L}]^2\rangle = \sum_{i,j} \langle \xi_i \xi_j \rangle
\sim N \int_0^{\infty} dr\, C(r) 
\ee
with $C(r)$ defined by $C(|i-j|)=\langle \xi_i \xi_j \rangle$.
Using Eq.~(\ref{eq:corr-SM}), we get
\be
\langle [\delta \mathcal{L}]^2\rangle \sim N e \int_0^{\infty} dr \, \exp(-r/\ell) \sim Ne\ell \sim N\Ted
\ee
thus recovering the second equality in Eq.~(10) of the main text.

\subsubsection*{Spring length distribution}

The probability $P(\boldsymbol \xi)$ is a multivariate Gaussian, so
that the marginal distribution $p(\xi) = \int d\xi_2\ldots
d\xi_N P(\boldsymbol \xi)$ is also a Gaussian:
\be
p(\xi) =
\frac{1}{\sigma(\Ted,\ms)\sqrt{2\pi}}
e^{-\frac{\xi^2}{2\sigma^2(\Ted,\ms)}}, \ee with \be
\sigma^2(\Ted,\ms) = (A^{-1})_{ii} = \frac{\ms
  \Ted}{\sqrt{2\Ted+\ms^2}}.
\ee
The ``mean-field'' entropy $s_{\rm mf}$ introduced in the text
can be computed as
\bea
s_{\rm mf} &=& - \int ~d\xi
~p(\xi) ~\ln[p(\xi)] \nonumber \\ &=& \frac{1}{2} +
\ln[\sigma(\Ted,\ms) \sqrt{2\pi}], \label{eq:marg_entr}
\eea
which yields the asymptotic result
\be
s_{\rm mf}(\Ted) \sim \ln \Ted \, , \qquad \Ted \rightarrow\infty \,.
\ee
The fact that the entropy of
Eq.~\ref{eq:marg_entr} in the infinite temperature limit diverges, and
is then wrong compared to Eq.10 in the Letter, is consistent with the
fact that formula in Eq.~\ref{eq:marg_entr} is approximation which
does not take into account the correlations.


\begin{thebibliography}{99}


\bibitem{EO89} 
S.~F. Edwards and R.~B.~S. Oakeshott, Physica A {\bf 157}, 1080 (1989).

\bibitem{ME89} 
A. Mehta and S.~F. Edwards, Physica A {\bf 157}, 1091 (1989).

\bibitem{EM94} 
S.~F. Edwards and C.~C. Mounfield,  Physica A {\bf 210}, 279 (1994);
Physica A {\bf 210}, 290 (1994).

\bibitem{EG98} 
S.~F. Edwards and D.~V. Grinev, Phys. Rev. E {\bf 58}, 4758 (1998).

\bibitem{BHDC15} 
D.~P. Bi, S. Henkes, K.~E. Daniels, and B. Chakraborty, Ann. Rev. Cond. Matt. Phys. {\bf 6}, 63 (2015).

\bibitem{BKVS00} 
A. Barrat, J. Kurchan, V. Loreto, and M. Sellitto, Phys. Rev. Lett. {\bf 85}, 5034 (2000). 



\bibitem{HHC07} 
S. Henkes, C.~S. O’Hern, and B. Chakraborty, Phys. Rev. Lett. {\bf 99}, 038002 (2007).

\bibitem{HC09} 
S. Henkes and B. Chakraborty, Phys. Rev. E {\bf 79}, 061301 (2009).

\bibitem{BE09} 
R. Blumenfeld and S.~F. Edwards, J. Phys. Chem. B {\bf 113}, 3981 (2009).

\bibitem{BJE12} 
R. Blumenfeld, J.~F. Jordan, and S.~F. Edwards, Phys. Rev. Lett. {\bf 109}, 238001 (2012).

\bibitem{BZBC13} 
D.~P. Bi, J. Zhang, R.~P. Behringer, and B. Chakraborty, Europhys. Lett. {\bf 102}, 34002 (2013).



\bibitem{NKBJN98} 
E.~R. Nowak, J.~B. Knight, E. Ben-Naim, H.~M. Jaeger, and S.~R. Nagel, Phys. Rev. E {\bf 57}, 1971 (1998).

\bibitem{SGS05} 
M. Schr\"oter, D.~I. Goldman, and H.~L. Swinney, Phys. Rev. E {\bf 71}, 030301(R).

\bibitem{LCDB06} 
F. Lechenault, F. da Cruz, O. Dauchot, and E. Bertin, J. Stat. Mech. P07009 (2006).

\bibitem{NRRCD09} 
S. McNamara, P. Richard, S. de Richter, G. Le Ca\"er, and R. Delannay, Phys. Rev. E {\bf 80}, 031301 (2009).




\bibitem{KM02} 
J. Kurchan and H. Makse, Nature {\bf 415}, 614 (2002).

\bibitem{M04} 
P.~T. Metzger, Phys. Rev. E {\bf 70}, 051303 (2004).

\bibitem{MD05} 
P.~T. Metzger and C.~M. Donahue, Phys. Rev. Lett. {\bf 94}, 148001 (2005).

\bibitem{PCN06} 
M. Pica Ciamarra, A. Coniglio, and M. Nicodemi, Phys. Rev. Lett. {\bf 97}, 158001 (2006)

\bibitem{BK15} 
V. Becker and K. Kassner, \emph{arXiv:1506.03288}.





\bibitem{BPS00} 
J.~J. Brey, A. Prados, and B. Sanchez-Rey, Physica A {\bf 275}, 310 (2000).

\bibitem{LD01} 
A Lef\`evre and D.~S. Dean, J. Phys. A {\bf 34}, L213 (2001).

\bibitem{L02} 
A. Lef\`evre, J. Phys. A {\bf 35}, 9037 (2002)

\bibitem{BFS02} 
J. Berg, S. Franz, and M. Sellitto, Eur. Phys. J. B {\bf 26}, 349 (2002).

\bibitem{DGL02} 
G. DeSmedt, C. Godr\`eche, and J.~M. Luck, Eur. Phys. J. B {\bf 27}, 363 (2002).

\bibitem{DGL03} 
G. DeSmedt, C. Godr\`eche, and J.~M. Luck, Eur. Phys. J. B {\bf 32}, 215 (2003).




\bibitem{CN00} 
A. Coniglio and M. Nicodemi, Physica A {\bf 296}, 451 (2001).

\bibitem{BKVS01} 
A. Barrat, J. Kurchan, V. Loreto, and M. Sellitto, Phys. Rev. E {\bf 63}, 051301 (2001). 

\bibitem{DL01} 
D.~S. Dean and A. Lef\`evre, Phys. Rev. E {\bf 64}, 046110 (2001).

\bibitem{LD03} 
A. Lef\`evre and D.~S. Dean, Phys. Rev. Lett. {\bf 90}, 198301 (2003).





\bibitem{BE03}  
R. Blumenfeld and S.~F. Edwards, Phys. Rev. Lett. {\bf 90}, 114303 (2003).

\bibitem{BE06} 
R. Blumenfeld and S.~F. Edwards, Eur. Phys. J. E {\bf 19}, 23 (2005).

\bibitem{BSWM08} 
C. Briscoe, C.~M. Song, P. Wang, and H.~A. Makse, Phys. Rev. Lett. {\bf 101}, 188001 (2008).

\bibitem{WSJM11} 
P. Wang, C.~M. Song, Y.~L. Jin, and H.~A. Makse, Physica A {\bf 390}, 427 (2011).

\bibitem{APF14} 
D. Asenjo, F. Paillusson, and D. Frenkel, Phys. Rev. Lett. {\bf 112}, 098002 (2014).


\bibitem{SL03} 
Y. Srebro and D. Levine, Phys. Rev. E {\bf 68}, 061301 (2003).






\bibitem{BK67}
R. Burridge and L. Knopoff, Bull. Seismol. Soc. Am. {\bf 57}, 341 (1967). 

\bibitem{CL89}
J.~M. Carlson and J.~S. Langer, Phys. Rev. A {\bf 40}, 6470 (1989).

\bibitem{GB10}
J.-C. G\'eminard and E. Bertin, Phys. Rev. E {\bf 82}, 056108 (2010).

\bibitem{BPG11}
B. Blanc, L.-A. Pugnaloni, and J.-C. G\'eminard, Phys. Rev. E {\bf 84}, 061303 (2011).

\bibitem{BGP14}
B. Blanc, J.-C. G\'eminard, and L.-A. Pugnaloni, Eur. Phys. J. E {\bf 37}, 112 (2014).


\bibitem{Langer}
J.~S. Langer, arXiv:1501.07228.



\bibitem{LeBellac}
M. Le Bellac, "Quantum and Statistical Field Theory", Oxford Science Publications (Oxford, 1992).




\end{thebibliography}

\begin{thebibliography}{99}

\bibitem{fonseca} 
C.M. da Fonseca, J. Petronilho, Numer. Math. {\bf 100}, 2005.


\end{thebibliography}
\end{document}